\documentclass[twocolumn]{svjour3}         
\smartqed  
\usepackage{graphicx}
\begin{document}

\title{Dark-energy constraints with ALMA polarization measurements
}
\subtitle{Synergies with CMB experiments}


\author{Paola Andreani         \and
        Carlo Baccigalupi 
}


\institute{P. Andreani\at
             European Southern Observatory,
             Karl Schwarzschild strasse 2
             85748 Garching, Germany \\
              \email{pandrean@eso.org}           
           \and
            C. Baccigalupi\at
              SISSA/ISAS, v. Beirut 2-4, 34014 Trieste, \& INFN, Sezione di Trieste, 
V. Valerio 2, 34127 Trieste, Italy\\ \email{bacci@sissa.it}
}
\date{Received: date / Accepted: date}
\maketitle

\begin{abstract}
The Cosmic Microwave Background (CMB) physics can be used to constrain the dark energy dynamics: B modes of the polarization
of the diffuse CMB emission as well as the polarized signal towards clusters of galaxies 
are sensitive to the Hubble expansion rate and thus the dark energy abundance in the early stage of cosmic acceleration. 
The first effect is sourced by gravitational lensing of large scale cosmic structures, the second is due to scattering
of the primary CMB temperature anisotropy quadrupole by free electrons in cluster plasma.
We are investigating the capabilities of ALMA to detect these effects and constrain the high-redshift dark energy abundance 
through measurements of sub-arcminute CMB anisotropies.

\end{abstract}


\section{Polarization induced by gravitational lensing of the CMB}
\label{pibglotcmb}

\begin{figure}
  \includegraphics[width=0.5\textwidth]{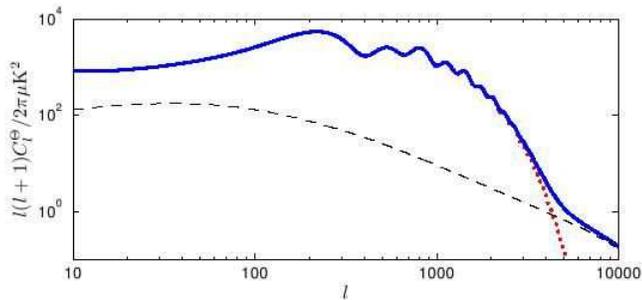}
\caption{
Expected signals: The lensed and unlensed TT polarization anisotropy spectrum C$_l^{TT}$, 
thick blue solid and red dotted lines, respectively. The dashed curve represents the small 
angle limit, which matches the curves on large multipoles \cite{LC}.}
\label{fig:1}       
\end{figure}

\begin{figure*}
  \includegraphics[width=0.85\textwidth]{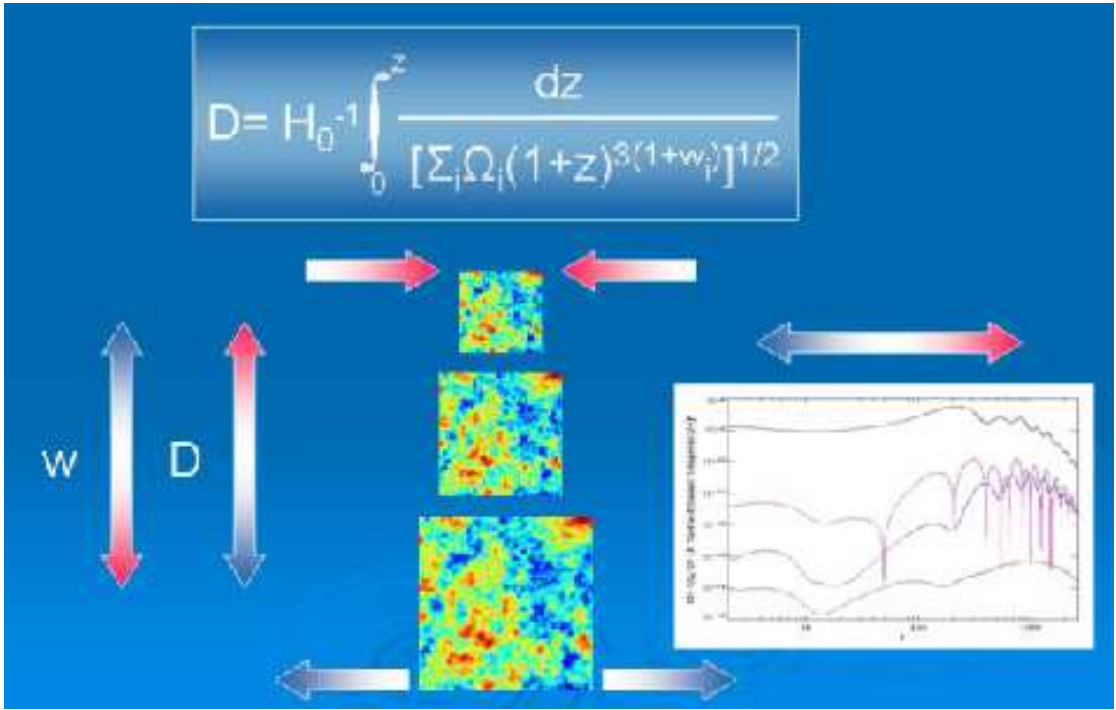}
\caption{The dark energy dynamics modifies the distance to last scattering: if the dark energy is more or less abundant
in the past with respect to now, the last scattering gets closer or 
farther re-projecting the CMB spots on larger or smaller angular scales, respectively. 
In the insert the power spectrum of the CMB anisotropies is shown. From the top: total intensity (TT), cross-correlation
between intensity and polarization (TE), gradient (EE) and curl (BB) power spectra of the CMB polarization. The spectra 
shift toward lower/higher multipoles l (larger/smaller angular scales) as dark energy increases/decreases with respect to 
the present abundance 
.}
\label{fig:2}       
\end{figure*}

The large scale density fluctuations in the Universe induce
deflections in the direction of the CMB photons 
as they propagate from the last scattering surface to the observer (see \cite{BS} for reviews). This gravitational 
lensing effect alters the anisotropy pattern of both the intrinsic temperature and polarization anisotropies. 
The net effect is a re-projection of the primary anisotropies at last scattering, which induces a correlation between 
different modes: because of gravitational lensing, the anisotropy power seen on a given angular scale is no longer 
due to a single cosmological mode, but rather to a set of those on a finite interval of scales which have been 
projected on that angle. \\
This has a number of effects on the CMB observables \cite{ZS}.
\begin{itemize}
\item
In total intensity, the lensing smears out the acoustic peaks in the power spectrum of CMB anisotropies, and adds power 
in the arcminute scale, where the primordial anisotropies decay naturally because of diffuse damping from Thomson 
scattering at decoupling. 
The relative changes $\delta C_{l}/C_{l}$, where $C_{l}$ represents the anisotropy 
power on a scale $l$ corresponding to $180/l$ degrees, is of the order $10\%$ on degree and tens arcminute 
scales, but diverges in the arcminute and sub-arcminute, as represented in Figure \ref{fig:1}. 
\item
In polarization, an additional effect occurs 
due to the leaking of the E modes into B ones. In the primary CMB anisotropies, the E modes are sourced by all 
kinds of cosmological perturbations, i.e. density, velocity and gravitational waves, while the B ones are excited 
by the latter two species, only. 
The lensing represents a second order effect in cosmology, having the cosmological perturbations in the large 
scale structures acting as lenses on the background CMB anisotropies. Along with the re-projection 
effect mentioned above, the lensing also mixes E and B modes (see the insert in Figure \ref{fig:2}). 
Since in the primordial perturbations 
one has E$>>$B, the net effect is perceived as a leaking of the E modes into 
B due to gravitational lensing \cite{ZS}. In the angular power spectrum, the B lensing modes peak at ten 
arcminute scale, corresponding to $l\simeq 1000$, resembling the primary E modes in shape, about $10$ times 
smaller. 
The lensing B signal is a fundamental prediction of the structure formation process, both in distribution 
of the angular power and amplitude, which may vary by a few ten percent level \cite{AB} depending on the 
dark energy abundance. 
\end{itemize}

Finally, the lensing affects the statistics of the CMB anisotropies, injecting a finite non-Gaussian 
power which is determined by the same correlation between scales. The non-Gaussian lensing power 
manifests itself in a bispectrum power which is predictable and may be connected with the most 
relevant cosmological parameters (see \cite{GBP} and references therein). 

\subsection{CMB lensing and dark energy}
\label{cmblade}

A very basic geometrical property of gravitational lensing is to be ineffective if the lens coincides 
with the source or the observer, respectively. In particular, if the CMB acts as a source, the lensing 
cross section peaks at $z\simeq 1$, with a width of about $0.5$. The latter epoch corresponds to the 
one in which the dark energy becomes dominating with respect to the non-relativistic matter.
The lensing of the CMB can be a probe of the dark energy abundance at the epoch of 
equality with matter. The lensing is sourced by cosmological structures, growing at a rate 
which is tightly related to the value of $H$, the Hubble expansion rate, at each time. The latter, through 
the Friedmann equation, is determined by the matter abundance, which is fixed at all times if its value is 
known today, and the dark energy one, which may vary according to the behavior of its equation of state. 
{\bf In summary, the CMB lensing probes $H$ at about $z=1\pm 0.5$, and so the dark energy abundance at the same 
epoch, being therefore complementary to the existing dark energy observables, mainly sensitive to its 
abundance at the present epoch.} 
It was shown \cite{AB} that within the present constraints on the value of the 
dark energy equation of state at the epoch of equality with matter, the amplitude of the 
BB modes presents variations of the order of $10\%$, which might be probed by the forthcoming 
experiments  
detecting the lensing distortion in the arcminute and sub-arcminute scale of CMB anisotropies.

\section{Polarization induced by the CMB quadrupole anisotropy}
\label{Pol}

Primary polarization is generated during the last scattering process and possesses a characteristic scale 
corresponding to the angle subtended by the horizon at recombination.
At later times, free electrons in the intergalactic medium of in clusters re-scatter CMB photons, 
causing secondary polarization. In the presence of dark energy, the gravitational potentials possess 
a time derivative which is different from zero due to the onset of acceleration, causing the 
Integrated Sachs-Wolfe (ISW) effect \cite{SW}. The latter is due to the net redshift or 
blueshift acquired by photons passing through evolving cosmological structures, i.e. collapsing 
over-densities or expanding voids as the photons cross them. 

Thus, the free electrons in clusters are hit by an anisotropic incident radiation, originated in part 
by the one present at last scattering, i.e. due to the primary anisotropies, and in part by the ISW effect. 
It has been shown \cite{SS} that the quadrupole component in the incident intensity 
distribution contributes to most of the polarization signal.\\
In addition, a local kinematic quadrupole arises 
from the cluster peculiar motion. The relevance of this process for dark energy is that the ISW quadrupole component 
is enhanced by the change in the cosmological equation of state due to the onset of acceleration. 

The primary quadrupole as well as the kinematic one represent unwanted contaminant for the effect induced 
by the ISW above. While the primary quadrupole is indistinguishable from the one induced by cosmological 
structures, the kinematic effect in clusters, due to the motion of free electrons with respect to the CMB frame, 
has a characteristic spectral dependence which provides with a potential method
to disentangle the two polarization effects.
    
\section{Observing polarization due to lensing on the CMB and clusters with ALMA}
\label{clusters}

\paragraph{clusters}
The cluster induced polarization can be derived by averaging the polarization signals taken 
from a large number of clusters located close to a given direction. 
The polarization effect depends linearly on the optical depth $\tau$ which has to be determined 
through an ensemble average. All other possible polarization effects associated with clusters are 
not additive so that the sum of the signals from individual clusters will tend to smooth out their contribution.

The polarization intensity depends on the frequency and therefore measurements in different bands must be carried out
and provide an opportunity to separate signals of different nature. Indeed, cosmologically induced cluster polarization 
signals can be affected by a strongly polarized cluster compact source. The polarization pattern in this case
is similar to that produced by the effect of finite optical depth, $\propto \eta \tau ^2$ \cite{SS} and its frequency dependence
(a power law) makes this signal easy to recognize. The unscattered polarization anisotropies, coming from the last scattering 
along the line of sight of a given cluster under consideration do not represent a significant contaminant, as they are 
damped on the angular scales of the cluster cores. Moreover, Faraday rotation measures of galaxy clusters are too small to 
produce a significant depolarization for wavelengths below 10 cm. On the other hand, 
high resolution observations are mandatory to single out the sources within the clusters and subtract them.

\paragraph{Polarization of the diffuse CMB emission}
On the angular scales probed by ALMA, the main foreground component for the CMB anisotropies in 
polarization is represented by extra-Galactic point sources. 
Polarization of the diffuse CMB emission can be estimated by averaging polarization signals taken 
from a large number of sky areas selected on the basis of their low foreground emission (very likely at high galactic latitudes). 
These regions will be observed at high spatial resolution in order to detect and subtract 
intense point sources and at the frequencies relevant for the CMB (1mm, 2mm, 800$\mu$m). The subtraction 
will contribute to reduce the effective noise caused by non-resolved point sources, when the same regions are 
observed with lower angular resolution and at the same frequencies in order to measure the 
CMB polarization anisotropies. 

\section{Polarization measurements}
\label{measurements}

\paragraph{Observing strategy}
Linear polarization measurements are possible at all ALMA bands.
Characterization of CMB polarization requires both very low noise and extreme control of systematic errors.
Interferometers offer several advantages: simple optics, instantaneous differencing of sky signals without scanning
and no differencing of detectors. The shape of the beam is well understood and the measurements are done directly in 
the Fourier space. Some sources of systematic error may cause the polarization signal to be contaminated by the much larger unpolarized
anisotropy, while 
others mix the E and B components. The operation of separating a polarization map into the E and B
components can be simplified if done in 
the Fourier or spherical harmonic space, i.e. it can be done locally (mode by mode), making the E-B separation cleaner\cite{SZ}.
Modeling systematic errors and instrumental noise is described in \cite{SZ} and \cite{DCJ}.
\\
The range of scales probed by an interferometer is determined by the spacing of the antennas, while the resolution in spatial
wave-numbers by the area of sky surveyed. By `mosaicing' several small patches of sky together a significant area of sky 
may be covered. The resolution in spatial wavenumber can be increased while the range
of spatial scales remain fixed by the geometry of the interferometer elements \cite{Bunn,BW}. 

\paragraph{Contamination from point sources}
The potential of polarized 
extra-Galactic sources to contaminate polarized CMB anisotropies was not well studied 
in the submm. 
Significant linear polarization is seen in most compact flat-spectrum radio sources which are the main contributors to small
scale foreground intensity fluctuations at $\lambda >$ 1 mm and the thermal dust emission from galaxies at submm $\lambda$
is also expected to be polarized to some extent. Average fractional polarization for BL Lac objects is 9 and 5 percent 
in the ALMA wavebands (i.e. \cite{ME,TC}). 
850$\mu$m polarization measurements towards Arp 220 put strong upper limits to its polarization degree of 1.54 per cent; 
very likely, polarized ULIRGs will not be a major contaminant \cite{SBS}.
Further studies in this field are mandatory.

\paragraph{Contamination from Galactic emission}
Galactic dust emission is observed to be polarized and 
the polarization increases with wavelengths, while its maximum 
decreases rapidly with increasing optical depth \cite{HIL}. 
The overall polarization level from the Galaxy
corresponds to the average of the contributions from regions with different polarizing efficiencies and different orientations 
of the magnetic field in the observed region of the sky, and the average polarization level decreases in comparison with individual 
clouds. Detection of diffuse polarized dust emission has been carried out by \cite{BE} and indicates a 3-5 \% 
polarization on large angular scales and low Galactic latitude. At higher latitudes the polarization fraction may increase due 
to the smaller averaging over different layers of the Galactic gas emitting in polarization. 
\hfill\break
A careful inspection of the 3-years WMAP sky indicates that at intermediate latitudes, in polarization the Galaxy dominates 
the sky signal at all frequencies \cite{CBC}. However, the power seems to be decaying on small angular scales, so that this 
is not expected to be a major contaminant for ALMA. Moreover, regions with low emission from Galactic synchrotron have 
been identified \cite{CBC}. On the other hand, the polarized thermal emission from Galactic dust grains is poorly known, 
and on the Galactic plane only \cite{BE}. 

%

\section{Conclusions}
We are investigating the capabilities of ALMA 
to detect the polarization of the CMB diffuse emission as well as 
the one from clusters of galaxies. The small field of view requires a mosaicing technique in order to map
the anisotropies on arcminute angular scales. In addition the very high resolution capabilities of ALMA will 
offer the unique opportunity to detect polarized sources in selected sky regions or in clusters. As these 
sources contaminate the CMB signals, ALMA should work in synergies with lower resolution CMB dedicated 
experiments to allow them to clean to reduce the contamination coming from unresolved point sources. 
%
%



\end{document}